\documentclass[review]{elsarticle}

\usepackage{lineno,hyperref,amssymb}
\modulolinenumbers[5]

\journal{J. Crystal Growth}

\begin{document}

\begin{frontmatter}

\date{November 13, 2019}

\title{Growth of CuFeO$_2$ Single Crystals by the Optical Floating-Zone Technique}

\author[1,2]{Nora Wolff\corref{mycorrespondingauthor}}
\cortext[mycorrespondingauthor]{Corresponding author}
\ead{nora.wolff@helmholtz-berlin.de}

\author[1]{Tobias Schwaigert}

\author[1]{Dietmar Siche}

\author[3,4]{Darrell G. Schlom}

\author[1]{and Detlef Klimm}

\address[1]{Leibniz-Institut f\"ur Kristallz\"uchtung, Max-Born-Str. 2, 12489 Berlin, Germany}

\address[2]{Helmholtz-Zentrum Berlin für Materialien und Energie, Hahn-Meitner-Platz 1, 14109 Berlin, Germany}

\address[3]{Department of Materials Science and Engineering, Cornell University, Ithaca, NY 14853-1501, USA}

\address[4]{Kavli Institute at Cornell for Nanoscale Science, Ithaca, NY 14853-1501, USA}

\begin{abstract}
CuFeO$_2$ single crystals up to 50\,mm in length and up to 10\,mm in diameter were grown by the optical floating-zone method. Stoichiometric polycrystalline rods with a diameter of 6--12\,mm were used as feed materials to produce crystals of sufficient size to be used as substrates for the growth of thin films of delafossites. For stable growth along the $c$-axis, low growth rates of 0.4\,mm/h are necessary. Due to the incongruent melting behavior of CuFeO$_2$, a stable melt zone requires adjustment of the lamp power during growth. The melting of CuFeO$_2$ is not simply incongruent because the thermodynamic equilibrium includes more than two solid phases and the melt; the gas phase is also involved. The crystals were characterized by X-ray diffraction and X-ray fluorescence measurements.
\end{abstract}


\end{frontmatter}



\section{Introduction}

While examining the mineralogical collection of the \'{E}cole Nationale des Mines in Paris, Charles Friedel found an artifact claimed to be ``Graphite from Catherinebourgh, Sibiria'' --- which instead he reported in the year 1873 to be composed from equimolar quantities of Cu$^+_2$O with the combination of Fe$^{3+}_2$O$_3$, about 3.5\% Al$_2$O$_3$ \cite{Friedel1873}. This chemical composition can be written as Cu(Fe,Al)O$_2$. The new mineral was given the name delafossite. In the years since Friedel's discovery a large number of other A$^+$B$^{3+}$O$_2$ compounds have become known that show basically the same structural features: BO$_6$ octahedra form layers that are stacked parallel to $(001)$, and these edge-sharing octahedral sheets are connected along the [001] direction by linear O\,---A$^+$---\,O bonds. Depending on details of the stacking sequence, the structures are usually either hexagonal or trigonal \cite{wolff_nora}. As an exception, Cu$^+$Mn$^{3+}$O$_2$ is monoclinic with space group $C2/m$ \cite{Topfer95}. Shannon et al. \cite{Shannon71} revealed that not only copper and silver can acts as the A$^+$ element, but also other quite noble metals like palladium or even pla\-tinum. This is surprising because oxides of the platinum group metals are not only scarce and often unstable, but also the known platinum group binary oxides show oxidation states of 2+ or higher, e.g. PdO, PdO$_2$, PtO, PtO$_2$, PtO$_3$, and not Pt$^{1+}$ as occours in Pt-containing delafossites.

Oxide materials based on the ABO$_2$ delafossite structure are of particular interest due to the novel properties that accompany their cation variation at A and B sites. These properties are of interest to fundamental science \cite{Hicks12, Takatsu07, Kushwaha15, Daou17, Sunko17} as well as applications \cite{Kawazoe00}. Usually, the semiconductor delafossites consist of Ag or Cu at the A-site and several trivalent cations like Al, Fe or Ga at the B-site. Important among the semiconducting delafossites is CuAlO$_2$ with its relatively high mobility for a p-type transparent conducting oxide \cite{Kawazoe97}. Pd- and Pt-based compounds (for A$^+$) are metallic delafossite oxides where B-site cations are transition metals like Co, Cr or Rh \cite{Shannon71, Prewitt71, Rogers71}. Among these, the growth of single crystalline PdCoO$_2$ has become of interest due to its ultra-high conductivity at room temperature. Recently an in-plane resistivity $\rho_{ab}=2.6\,\mu\Omega\cdot$cm at 295\,K was measured for sub-mm-sized PdCoO$_2$ crystals, which makes this material the most conductive oxide known, comparable to the best metallic conductors Ag, Cu, Au and Al \cite{Hicks12}. Even though it is a platinum group oxide, this material has a fairly high thermal stability up to $900^{\,\circ}$C \cite{Shannon71}.

The excellent properties exhibited by delafossites, including a mean-free path of about 20$\mu$m at low temperature in PdCoO$_2$, \cite{Hicks12} invites the growth of delafossite heterostructures. In such heterostructures bandgap engineering \cite{Capasso87}, strain engineering \cite{Schlom14}, symmetry breaking and other thin-film approaches could be applied to modify the properties of delafossites as has become commonplace for other oxides \cite{Ramesh19}. Unfortunately, there are no commercially available delafossite substrates.

So far, epitaxial layers have mainly been grown on sapphire substrates \cite{Brahlek19, Harada18, Yordanov19} -- unfortunately with significantly degraded quality resulting from the not well matched crystallographic lattice of sapphire ($R\bar{3}c$, $a=4.7602$\,\AA, $c=12.9933$\,\AA\ \cite{Lewis82}) and PdCoO$_2$ ($R\bar{3}m$ $a=2.830$\,\AA, $c=17.743$\,\AA\ \cite{Takatsu07}). The resulting delafossite films contain in-plane rotation twins and other defects that result in the best of today's PdCoO$_2$ films \cite{Brahlek19} having resistivities at room temperature 1.8 times higher than PdCoO$_2$ single crystals \cite{Mackenzie17} and 145 times higher resistivities at 4 K. For the growth of high quality epitaxial layers, isostructural single crystalline substrates with similar lattice constants are required. 

The aim of this work is to grow relatively large delafossite crystals that in subsequent studies can be used as substrates for the growth of high quality delafossite films. According to the structural parameters, CuAlO$_2$ ($R\bar{3}m$, $a=2.8571$\,\AA, $c=16.940$\,\AA\ \cite{Shannon71}) would be the most suitable substrate material for PdCoO$_2$. Unfortunately, the melt growth of bulk CuAlO$_2$ crystals with sufficient size is impossible, because the substance melts peritectically under the formation of solid Al$_2$O$_3$ and the growth window (= CuAlO$_2$ liquidus) between the peritectic and eutectic points is too small \cite{Wolff18,Wolff19}. The melting points of iron oxides (Fe$_2$O$_3$ decomposes to Fe$_3$O$_4$, which melts at $1597^{\,\circ}$C) are significantly lower than for Al$_2$O$_3$ ($T_\mathrm{f}=2054^{\,\circ}$C \cite{FactSage7_4}), and consequently a larger growth window for CuFeO$_2$ ($R\bar{3}m$, $a=3.0351$\,\AA, $c=17.166$\,\AA) can be expected. (Some CuO--Fe$_2$O$_3$ phase diagrams \cite{Zhao95b,Song16} in the literature neglect the decomposition of Fe$_2$O$_3$ and show the congruent melting point of Fe$_3$O$_4$ instead.) CuFeO$_2$ is the only delafossite compound, that has already be grown with reasonable quality in diameters up to 5--8\,mm by the optical floating zone (OFZ) technique \cite{Zhao95b,Zhao95,Zhao96}.

So far CuFeO$_2$ single crystals have been grown from stoichiometric sintered rods by the OFZ technique. Our prior thermodynamic investigations have shown that growth methods involving crucibles are unsuitable as the aggressive copper melt attacks all crucible materials \cite{Wolff18}. In addition to pure CuFeO$_2$, CuFe$_{1-x}$Ga$_x$O$_2$ with $x\leq\ $12\% have been grown by Song et al. by the OFZ method \cite{Song16}. For CuCr$_{1-x}$Al$_x$O$_2$ it was shown that solid solution formation with mixed occupancy on the B-site seems to stabilize the delafossite phase thermodynamically \cite{Kato17}, but CuFeO$_2$ with high ($\approx30$\%) Al$^{3+}$ content only grew in polycrystalline form, presumably as a result of severe segregation in the CuFeO$_2$--CuAlO$_2$ solid solution system. The present work focuses on the growth of pure CuFeO$_2$ with sufficient size and perfection to be useful for delafossite substrates. We show that the oxygen fugacity $p_{\mathrm{O}_2}$ in the gas phase acts on the equilibria between the relevant oxidation states of copper ($+2,+1,0$) and iron ($+3,+2$) --- which then influences the crystallization behavior of delafossites.


\section{Experimental}

Growth experiments were carried out in an OFZ furnace from Crystal System Corporation (type FZ-T-10000-H-VII-VPO-PC). The equipment has a four-mirror setup with 1000\,W or 1500\,W halogen lamps and a fused silica protection tube. An argon flow of 0.3\,l\,min$^{-1}$ with 5N purity and total pressure of 1 atm was used as the growth atmosphere, to stabilize the CuFeO$_2$ phase \cite{Wolff19}. The seed and feed rods were rotated at 15\,rpm in opposite directions, and pulling rates between 0.4 and 1.1\,mm\,h$^{-1}$ were employed. Due to the incongruent melting, the lamp power was decreased slowly during the first 5-10\,h of the experiments.

To prepare the rods, vacuum-dried powders of Cu$_2$O (Fox Chemicals, 4N purity) and Fe$_2$O$_3$ (Alfa Aesar, 4N8 purity) in 1:1 molar ratio were mixed and sintered for 25\,h at $900^{\,\circ}$C in 1 atm of argon. XRD powder analysis confirmed that a pure delafossite phase resulted from this annealing process. Following grinding of the sintered CuFeO$_2$, the powder was compacted in a plastic bag and pressed for 3\,min at 2\,kbar at room temperature in a cold isostatic press from Engineered Pressure International NV, Belgium. To increase the density, the pressed block was again sintered for 8\,h at $900^{\,\circ}$C in argon. Rods with diameters of 6, 8 and 12\,mm and 120\,mm length were drilled out of the sintered blocks with a core drill. The density of the rods was 70-80\% of the theoretical value. The grown crystals were characterized by X-ray diffraction (XRD), energy-dispersive Laue mappings (EDLM) and X-ray fluorescence (XRF) spectroscopy. The latter were performed using a $\mu$-XRF spectrometer M4 Tornado (Bruker). Laue method was used to orient the crystals and prepare planar substrates oriented perpendicular to the $c$-axis.


\section{Results}

Five growth experiments were performed that are summarized in Table~\ref{tab:OZF}. In order to optimize the quality and size of the crystals, the diameter of the polycrystalline source rods was increased and the growth rate was reduced in the later growth experiments. Due to the incongruent melting of the compound, slow growth rates $<0.4$\,mm\,h$^{-1}$ are preferable and gave the best results. In addition, lower growth rates facilitate the transport of oxygen gas from the environment to the growth interface to oxidize the iron (Fe$^{2+}\rightarrow$~Fe$^{3+}$), as is discussed in the next section. Higher growth rates lead to polycrystalline growth, and the crystal orientation during the growth experiment often changes. At low growth rate the $c$ axis is the prevailing growth direction. Occasional changes in growth orientation are clearly visible on the crystal surface of the grown rod because the $(001)$ planes of the delafossite are shiny and the others appear dull. This means that a crystal that is grown along $\left[001\right]$ should not have a shiny lateral surface. In contrast to recent observations during the growth of CuLaO$_2$ delafossite in pure argon \cite{Mohan14}, no outer ring containing Cu$^{2+}$ was observed in our experiments with CuFeO$_2$.

\begin{figure}[htb]
\includegraphics[width=0.70\textwidth]{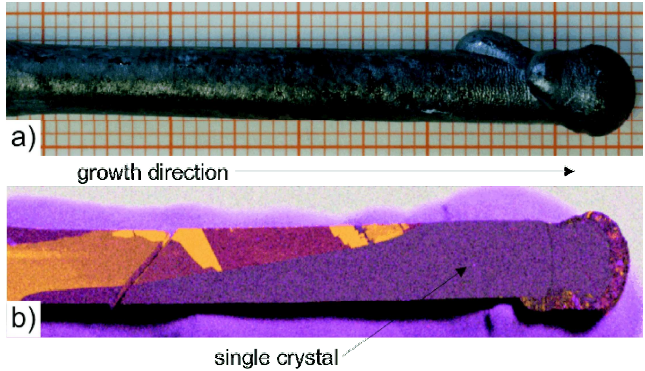}
\caption{a) CuFeO$_2$ grown rod \#1 (cf. Table~\ref{tab:OZF}). b) EDLM results of the same rod sectioned parallel to the growth direction.}
\label{fig:Xtal-1}
\end{figure}

Fig.~\ref{fig:Xtal-1} shows a rod that was grown with a high growth rate of 1.1\,mm\,h$^{-1}$. In order to visualize the single crystalline regions and composition, the rod was sawn along its axis and investigated by EDLM. Elemental maps of the cross sections revealed a composition close to the stoichiometric of the grown material. Spatial mapping of a Bragg peak (EDLM), showed that only the tail end of the grown rod is a single crystal \cite{Guguschev15a}.

\begin{table}[ht]
\centering
\caption{OFZ growths experiments with CuFeO$_2$. $d$ -- rod diameter, $v$ -- growth rate, $P$ -- lamp power.}
\begin{tabular}{lcccl}
\hline
Grown Rod \# & $d$ (mm) & $v$ (mm\,h$^{-1}$) & $P$ (W) & remarks \\ \hline
1             & 6        & 1.1                & 1000    & Fig.~\ref{fig:Xtal-1} \\ 
2             & 8        & 0.55               & 1000    & Fig.~\ref{fig:Xtal-3}a)       \\ 
3             & 8        & 0.4                & 1000    & Fig.~\ref{fig:Xtal-3}b) \\ 
4 \& 5        & 12       & 0.4                & 1500    & Fig.~\ref{fig:Xtal-4}, best results \\ \hline
\end{tabular}
\label{tab:OZF}
\end{table}

\begin{figure}[htb]
\includegraphics[width=0.70\textwidth]{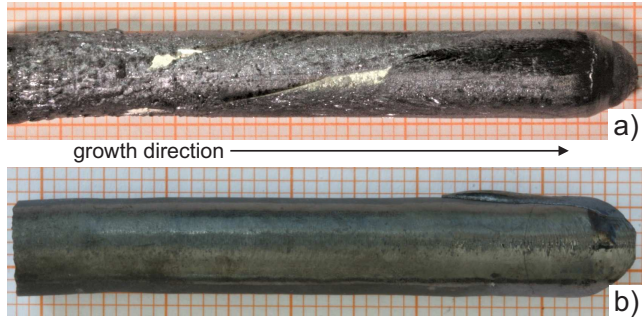}
\caption{a) CuFeO$_2$ rod \#2. b) CuFeO$_2$ rod \#3 (cf. Table~\ref{tab:OZF}).}
\label{fig:Xtal-3}
\end{figure}

The reduction of the growth rate to 0.55\,mm\,h$^{-1}$ and then to 0.4\,mm\,h$^{-1}$ resulted in significantly better crystal quality (Fig.~\ref{fig:Xtal-3}a) and b), respectively). With a growth rate of 0.4\,mm\,h$^{-1}$, the change in crystal orientation during the growth process could be avoided. Even though the $c$-axis is slightly tilted with respect to the growth direction (part of the lateral surface is shiny), the longest delafossite single crystal known to date was grown.

\begin{figure}[htb]
\includegraphics[width=0.70\textwidth]{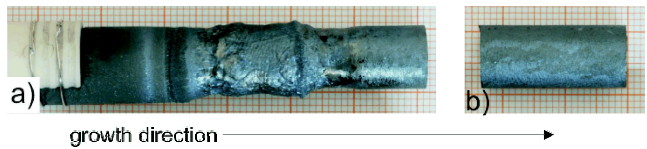}
\caption{a) Unstable growth of CuFeO$_2$ in the initial growth state of rod \#5 (cf. Table~\ref{tab:OZF}). b) Single crystal part of the same rod.}
\label{fig:Xtal-4}
\end{figure}

The handling of larger diameters $\geq10$\,mm is more difficult because the surface tension of the larger melt zone is less stable, resulting in overflow (Fig.~\ref{fig:Xtal-4}a)). Stable growth became possible if the diameter of the growing rod did not exceed 10\,mm. Fig.~\ref{fig:Xtal-4}b) shows the separated single crystalline part of the grown rod, which is homogeneous without any shiny $(001)$ planes on the surface. This part has a size of 10\,mm in diameter and 23\,mm in length. Laue measurements at both cutting ends confirmed the segment of the rod to be a single crystal grown along the $c$-axis without any tilting. The crystal that was grown at a rate of 0.4\,mm\,h$^{-1}$ will be used as substrate material for the growth of PdCoO$_2$ films by molecular-beam epitaxy.

\section{Discussion}

Like other delafossites, CuFeO$_2$ is known to melt incongruently. This means the solid (with Cu:Fe = 1:1) is not in equilibrium with a liquid of the same composition. Typically the melting behavior is described to be peritectic, which means that upon heating in addition to melt with different composition some higher melting solid phase is formed. For CuAlO$_2$, $\alpha$-Al$_2$O$_3$ is the higher melting phase, and in some recent publications Fe$_2$O$_3$ is claimed to play this role for CuFeO$_2$ \cite{Zhao95b,Song16}. This assumption is, however, incorrect because Fe$_2$O$_3$ decomposes to Fe$_3$O$_4$ and cannot be treated as a stable component of the system. Instead Cu, Fe and O$_2$ should be chosen as system components, see e.g. Fig.~1 in Ref. \cite{Wolff19}.

\begin{figure}[htb]
\includegraphics[width=0.49\textwidth]{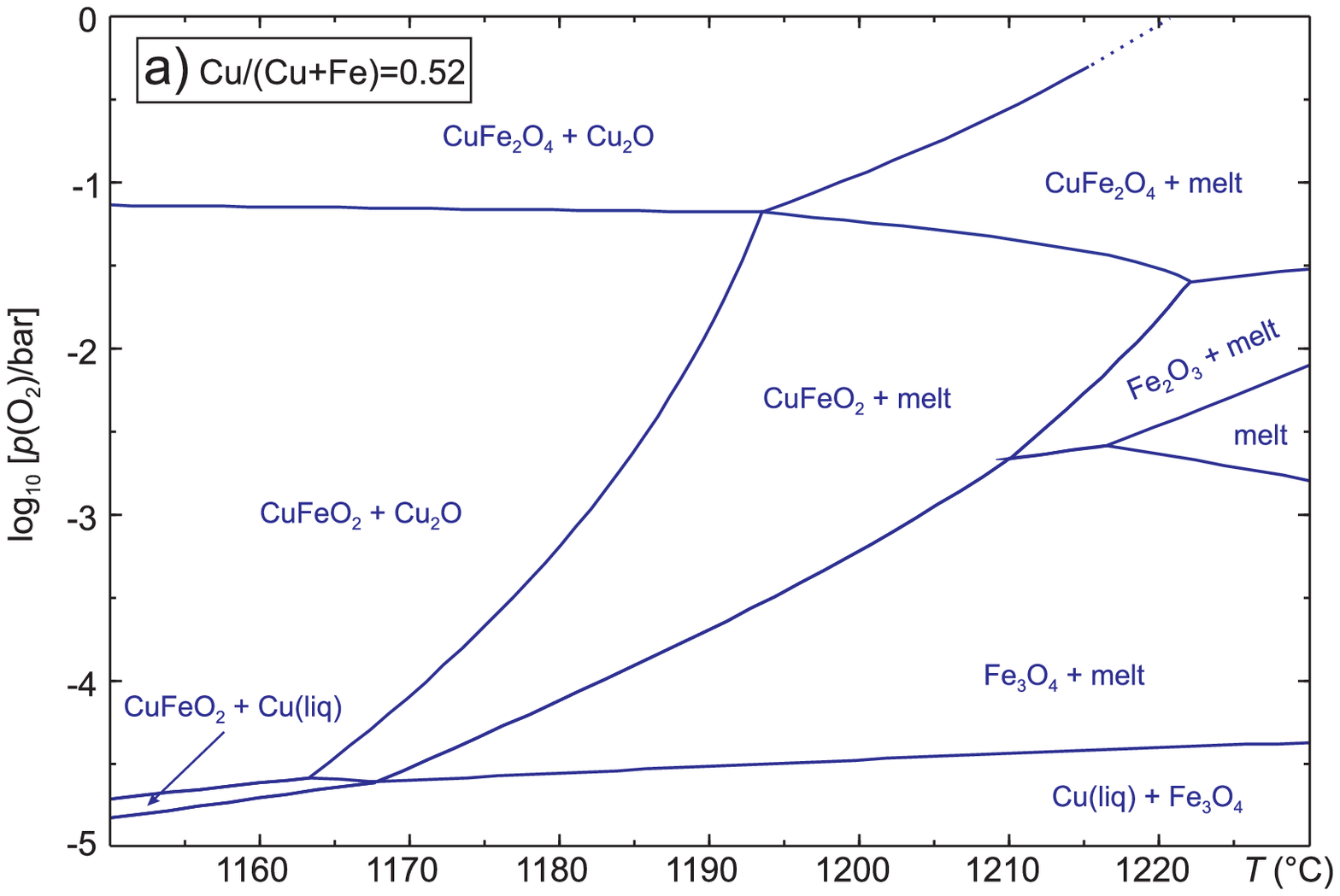}
\includegraphics[width=0.49\textwidth]{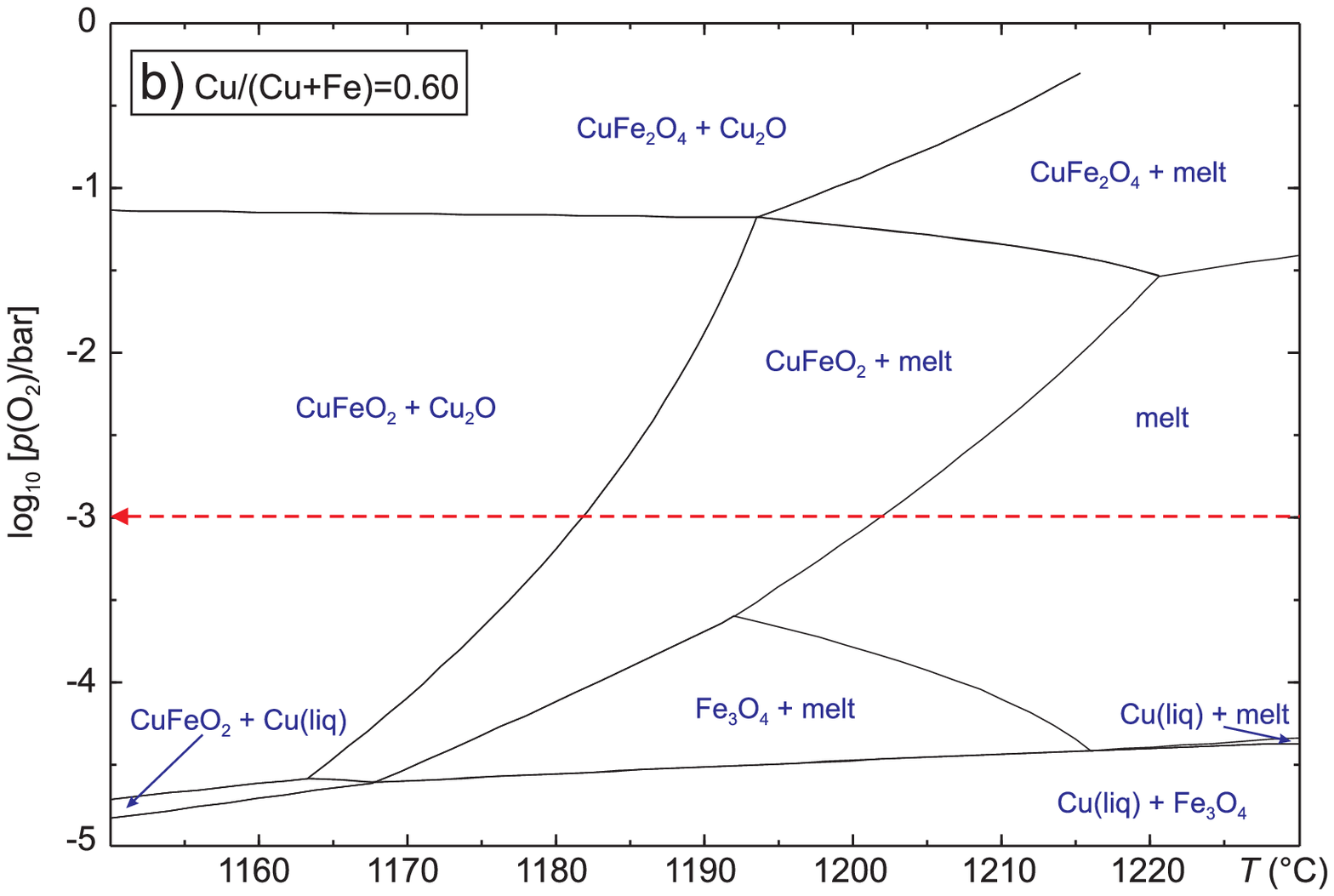}
\caption{Predominance phase diagram of the system Cu--Fe--O$_2$ for a molar Cu concentration of a) 52\%; b) 60\%. The red (dashed) arrow at $\log_{10}[p(\mathrm{O}_2)/\mathrm{bar}]=-3$ is explained in Fig.~\ref{fig:cooling_06}.}
\label{fig:PDs}
\end{figure}

More detailed FactSage \cite{FactSage7_4} calculations of the melting behavior of the Cu--Fe--O system are shown in Fig.~\ref{fig:PDs}. In the left phase diagram a fixed copper concentration of 52\,mol-\% is assumed, which is only a small Cu excess over Fe. On the right rim a ``melt'' phase field exists were the whole system forms a single copper-iron-oxide melt around $\log_{10}[p(\mathrm{O}_2)/\mathrm{bar}]\approx-2.5$. Upon cooling from this melt either Fe$_2$O$_3$ or Fe$_3$O$_4$ crystallizes first, depending on the oxygen fugacity. (For high $p_{\mathrm{O}_2}$ or high Cu excess the spinel Cu$^+$Fe$^{3+}_2$O$_4$ can also crystallize first.) The primary crystallization of iron oxide from Cu:Fe=1:1 mixtures was also observed experimentally (Fig.~\ref{fig:Xtal-1}) and depletes the melt of iron. 

A thermodynamic assessment of the complete Cu--Fe--O system is beyond the scope of this paper. Such assessments have been provided by Khvan et al. \cite{Khvan11} and more recently by Shishin et al. \cite{Shishin13}. Irrespective of differences in the models that are used by these authors, they agree on several key points that are relevant for the current crystal growth experiments. First, at temperatures between 1100--1400 K the delafossite CuFeO$_2$ is stable only for medium oxygen fugacities $\approx10^{-7}-10^{-1}$\,bar, which overlaps well with the experimental conditions of this study. Second, the melt consists basically of Cu$^+$, Cu$^{2+}$, Fe$^{2+}$, Fe$^{3+}$ and O$^{2-}$. Minor amounts of Cu$^0$ and Fe$^0$ are present for metal-rich melts. Cu$^{3+}$, which was reported under very high oxygen pressure \cite{Schramm05}, can be neglected under the growth conditions we use in this study. 

The authors of reference \cite{Shishin13} include several of the authors of the FactSage \cite{FactSage7_4} thermodynamic system, and consequently the FactSage calculations that are used in the present paper are in good agreement with both prior assessments \cite{Khvan11,Shishin13} --- except the circumstance that FactSage so far contains no data for Cu$^{2+}$ (CuO) in the melt. Nevertheless, under the growth conditions for CuFeO$_2$ the CuO content was found to be low, $<10$\% \cite{wolff_nora}, which agrees well with the literature (see Fig.~16 in Ref. \cite{Shishin13}). Consequently, these current FactSage calculations give a realistic description of the growth experiments.

The ``CuFeO$_2$+melt'' phase field has no connection to ``melt'' because the delafossite melts peritectically. Depletion of the melt by iron, however, shifts the Cu:Fe ratio towards larger values. In Fig.~\ref{fig:PDs}b) the Cu concentration is increased to 60\,mol-\% and this higher copper concentration leads to significantly lower liquidus temperatures. Between $1220^{\,\circ}$C and $1192^{\,\circ}$C both phase fields are directly connected and along this curve CuFeO$_2$ crystallizes first. In typical binary concentration vs. temperature ($x-T$) phase diagrams peritectics are horizontal (isotherm) lines in the diagram; this is not so here where $p(\mathrm{O}_2)$ influences the temperature where the solid crystallizes.

\begin{figure}[htb]
\includegraphics[width=0.70\textwidth]{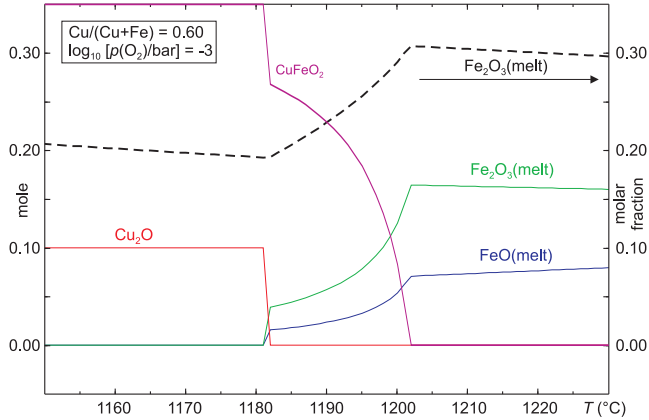}
\caption{Crystallization of the Cu-Fe-oxide melt that is shown in Fig.~\ref{fig:PDs}b) under a constant oxygen fugacity of 1\,mbar: Even if only CuFe$^{3+}$O$_2$ starts to crystallize around $1200^{\,\circ}$C, the amount of Fe$^{2+}$O in the melt drops.}
\label{fig:cooling_06}
\end{figure}

Fig.~\ref{fig:cooling_06} (which should be read from the right to the left) analyzes the crystallization path that is marked by a red dashed arrow in Fig.~\ref{fig:PDs}b). For the sake of simplicity, only the relative amounts (in moles) of Fe(II) oxide and Fe(III) oxide in the melt, the corresponding molar fraction of Fe$_2$O$_3$(melt), and the amounts (in moles) of solid CuFeO$_2$ and Cu$_2$O are shown. It is evident that above the $1202^{\,\circ}$C liquidus the amount of Fe$_2$O$_3$(melt) rises slightly at the expense of FeO(melt) because a lower temperature shifts the equilibrium to higher valency. The first solid to crystallize is thus CuFeO$_2$, which contains only Fe$^{3+}$. With cooling below $1202^{\,\circ}$C, not only does the fraction of Fe$_2$O$_3$(melt) decrease, but so does FeO(melt). Obviously additional free oxygen must be absorbed from the gas phase to oxidize FeO to Fe$_2$O$_3$, from which the delafossite finally forms. After passing the left boundary of the ``CuFeO$_2$+melt'' phase field at $1081^{\,\circ}$C, the melt disappears completely and solid Cu$_2$O + additional CuFeO$_2$ crystallize together. This corresponds to the eutectic point in a standard $x-T$ phase diagram.

Note that the oxygen fugacity $\log_{10}[p(\mathrm{O}_2)/\mathrm{bar}]=-3$ used for the calculation of Fig.~\ref{fig:cooling_06} is expected to be close to the true conditions used in the crystal growth experiments. In the ideal case, one can assume that 5N (99.999\%) Ar contains oxygen background impurities around $2\times10^{-6}$\,bar, which sets a lower boundary on the oxygen partial pressure used during growth \cite{Klimm05b}. The leaks in the growth chamber and out-gassing of chemicals and constructed parts will further increase the oxygen level. But even if the estimated oxygen fugacity is incorrect by one or two orders of magnitude, or if the database and our own thermodynamic values \cite{wolff_nora,FactSage7_4} that were used for the calculation of Figs.~\ref{fig:PDs} and \ref{fig:cooling_06} would be somewhat erroneous, the general topology is not significantly changed. Hence, these figures are expected to give a realistic model.

Copper can exist in the melt with oxidation states $+2,+1$ and $0$, as shown in Fig.~6 of \cite{Wolff18}, but with recent thermodynamic data for copper oxide melts \cite{wolff_nora} one finds that, under the conditions of CuFeO$_2$ growth, $>90$\% of the copper atoms are Cu$^+$, which is the same valency as in the solid.

The importance of the gas phase to the crystal growth process is not restricted to the delafossite CuFeO$_2$; during the growth of the delafossite CuAlO$_2$, the gas phase is also significantly involved. Fig.~4 in our recent paper \cite{Wolff18} shows DTA/TG heating curves for several Cu$_2$O--Al$_2$O$_3$ mixtures, and the peritectic melting of CuAlO$_2$ can clearly be seen as an endothermic peak near $1230-1250^{\,\circ}$C for the samples with 6\% and 10\% Al$_2$O$_3$. The melting peak is always accompanied by a small mass gain. If Cu$^+$Al$^{3+}$O$_2$ melts, Al$^{3+}$ maintains its oxidation state. Cu$^+$ is partially oxidized to Cu$^{2+}$ because a deep eutectic (ca. 150\,K below the congruent melting point of pure Cu$_2$O) appears between Cu$_2$O and CuO where the melt is entropically stabilized \cite{Schramm05}. This means that upon cooling to achieve crystal growth, CuO must be partially reduced to Cu$_2$O, releasing free oxygen. Beyond the unfavorable position of the peritectic and eutectic points in this system, the permanent production of free oxygen at the phase boundary is detrimental to crystal growth. This is in contrast to the Cu--Fe--O system described here, because the necessary oxidation of Fe$^{2+}$ to Fe$^{3+}$ can be performed without the formation of gas. For the growth of CuFeO$_2$ the oxygen production that would result from Cu$^{2+}\rightarrow$\,Cu$^+$ reduction is likely compensated by the Fe$^{2+}\rightarrow$\,Fe$^{3+}$ oxidation. Indeed, during DTA/TG experiments, no mass change is observed during melting or crystallization, because the O$_2$ exchange with the atmosphere is insignificant. Another potential explanation for the lack of mass change could be the partial incorporation of Cu$^{2+}$ into the delafossite structure, which was observed e.g., for LaCuO$_{2.5+x}$ and YCuO$_{2.5+x}$ delafossites \cite{Cava93}. A significantly different oxidation behavior of Cu$^+$B$^{3+}$O$_2$ delafossites for different B$^{3+}$ ions was reported elsewhere \cite{Amrute13}.


\section{Conclusions}

Using the OFZ technique single crystals of CuFeO$_2$ with diameters up to 10\,mm are grown. These are the largest synthetic delafossite single crystals ever grown and have sufficient size and quality for the preparation of wafers for the epitaxial deposition of thin films of other functional delafossites. The growth is performed from stoichiometric rods prepared from 1:1 (molar) mixtures of Cu$_2$O and Fe$_2$O$_3$ in an argon atmosphere. The melting behavior of CuFeO$_2$ is almost peritectic, by which we mean that the material CuFeO$_2$ crystallizes from a melt that is enriched in Cu$_2$O/CuO. A low but significant oxygen fugacity ($\approx1$\,mbar) must be available in the growth atmosphere to oxidize Fe$^{2+}$, which is partially present in the melt, to Fe$^{3+}$ which is incorporated into the growing CuFeO$_2$ crystal. Hence, the melting behavior of CuFeO$_2$ is not simply peritectic (which would mean that two solid phases are in equilibrium with the melt). Rather the gas phase is also involved in the crystallization process, which makes crystal growth more complex and requires comparably low growth rates (cf. Table~\ref{tab:OZF}).


\section*{Acknowledgments}

A. Kwasniewski is acknowleged for performing X-ray powder analysis and Laue measurements. We are indebted to C. Guguschev who performed energy-dispersive Laue mappings and X-ray fluorescence spectroscopy and to S. Ganschow for continuous support of this work. The authors acknowledge funding by the German Research Foundation (DFG) under project SI 463/9-1. D.G.S. acknowledges funding provided by the Alexander von Humboldt Foundation for his sabbatical stay at the Leibniz-Institut für Kristallzüchtung and support from the U.S. Department of Energy, Office of Basic Sciences, Division of Material Sciences and Engineering, under Award No. DE-SC0002334.


\end{document}